\documentclass[aps,prl,reprint,superscriptaddress,nofootinbib,floatfix,longbibliography]{revtex4-2}
\usepackage[dvipsnames]{xcolor}
\usepackage{amssymb,amsmath,amsfonts,bm,slashed,soul,ragged2e,graphicx,epstopdf,hyperref,array}
\usepackage[utf8]{inputenc}
\usepackage{colortbl,color}
\usepackage{cancel}
\definecolor{steelblue}{RGB}{25,25,112}
\definecolor{dullblue}{rgb}{0,0.298,0.49}
\definecolor{darkred}{rgb}{0.545,0,0}
\definecolor{darkorange}{RGB}{222,132,69}
\definecolor{darkgreen}{RGB}{126,171,85}
\definecolor{blue2}{cmyk}{1, 0.1, 0.1, 0}
\hypersetup{colorlinks,linkcolor={darkred},citecolor={blue},urlcolor={dullblue}}

\usepackage{comment}
\usepackage{braket}
\usepackage{soul}
\usepackage{makecell}
\usepackage{amsmath}

\newcommand{\beq}{\begin{equation}}
\newcommand{\eeq}{\end{equation}}
\newcommand{\bea}{\begin{eqnarray}}
\newcommand{\eea}{\end{eqnarray}}

\newcommand{\gsim}{\lower.7ex\hbox{$\;\stackrel{\textstyle>}{\sim}\;$}}
\newcommand{\lsim}{\lower.7ex\hbox{$\;\stackrel{\textstyle<}{\sim}\;$}}

\newcommand{\be}{\begin{equation}}
\newcommand{\ee}{\end{equation}}
\newcommand{\ba}{\begin{eqnarray}}
\newcommand{\ea}{\end{eqnarray}}

\newcommand{\D}{\mathrm{d}}

\newcommand{\e}{\mathrm{e}}

\definecolor{chggreen}{RGB}{0,100,0}

\RequirePackage[normalem]{ulem}

\newcommand{\unipd}{Dipartimento di Fisica e Astronomia ``G. Galilei'', Università degli Studi di Padova, via Marzolo 8, I-35131 Padova, Italy}
\newcommand{\infnpd}{INFN, Sezione di Padova, via Marzolo 8, I-35131 Padova, Italy}
\newcommand{\tdli}{State Key Laboratory of Dark Matter Physics, Tsung-Dao Lee Institute and School of Physics and Astronomy,\\ Shanghai Jiao Tong University, Shanghai 201210, China}
\newcommand{\sjtu}{Key Laboratory for Particle Astrophysics and Cosmology (MOE), and Shanghai Key Laboratory for Particle Physics and Cosmology, Shanghai Jiao Tong University, Shanghai 201210, China}

\usepackage{orcidlink}

\begin{document}

\title{Audible curvaton: modulated phase transition as the origin of large-scale curvature perturbations and gravitational wave anisotropies}

\author{Wen-Yuan Ai\,\orcidlink{0000-0002-6042-7407}}
\email{wenyuanai@sjtu.edu.cn}
\affiliation{\tdli}
\affiliation{\sjtu}

\author{Gabriele Franciolini\,\orcidlink{0000-0002-6892-9145}}
\email{gabriele.franciolini@unipd.it}
\affiliation{\unipd}
\affiliation{\infnpd}

\author{Xin Wang\,\orcidlink{0000-0003-4292-460X}}
\email{xin.wang@unipd.it}
\affiliation{\unipd}
\affiliation{\infnpd}

\author{Xiang-Xi Zeng\,\orcidlink{0009-0003-0632-3230}}
\email{zengxiangxi@sjtu.edu.cn}
\affiliation{\tdli}
\affiliation{\sjtu}

\begin{abstract}

We propose a post-inflationary first-order phase transition modulated by a light spectator field as the origin of primordial large-scale curvature perturbations. Spectator fluctuations acquired during inflation modulate the bubble nucleation history, generating curvature perturbations together with correlated intrinsic gravitational-wave (GW) anisotropies. We show that this mechanism can reproduce the observed amplitude of curvature perturbations while producing a loud stochastic GW background in the frequency bands of space-based detectors such as LISA, Taiji, and TianQin. A concrete benchmark yields strongly enhanced intrinsic GW anisotropies and a local primordial non-Gaussianity signal within reach of future large-scale-structure surveys.

\end{abstract}

\date{\today}

\maketitle

\textbf{\textit{Introduction.}}---The observations of the cosmic microwave background (CMB) have established a nearly scale-invariant spectrum of primordial curvature perturbations~\cite{Planck:2018jri}. Inflation~\cite{Starobinsky:1980te,Guth:1980zm,Linde:1981mu,Albrecht:1982wi} provides a compelling framework for their origin, but the perturbations need not originate from the inflaton itself. A light spectator field can acquire fluctuations during inflation and later transfer them to radiation. In the conventional curvaton mechanism, the spectator oscillates after inflation, grows in relative importance, and transfers its perturbations through its decay~\cite{Linde:1996gt,Enqvist:2001zp,Lyth:2001nq,Moroi:2001ct,Bartolo:2002vf}.

The essential ingredient is the conversion of initially isocurvature perturbations into adiabatic curvature perturbations after inflation. The same principle underlies modulated reheating~\cite{Dvali:2003em,Kofman:2003nx,Matarrese:2003tk}, in which a light spectator spatially modulates the inflaton decay rate. The inflaton supplies the energy released into radiation, while the spectator controls the reheating time and can remain energetically negligible.

In this Letter, we propose a curvaton-modulated first-order phase transition (FOPT) to realize this conversion.
FOPTs can generate small-scale curvature perturbations at the time of the phase transition (PT)~\cite{Liu:2022lvz,Lewicki:2024ghw,Franciolini:2025ztf,Sui:2025epg,Li:2026hef} and can also probe primordial isocurvature through the generation of anisotropic gravitational waves (GWs)~\cite{Geller:2018mwu,Kumar:2021ffi,Bodas:2022urf,Cui:2023dlo,Li:2021iva}. By contrast, we consider a post-inflationary FOPT whose modulation by a light spectator field generates {\it large-scale} curvature perturbations.\footnote{Modulated continuous PTs have been studied previously~\cite{Matsuda:2009yt,Dimopoulos:2012nj} but have received relatively little attention. FOPTs triggered or modulated by an additional field have also been studied in the context of inflation~\cite{Buckley:2024nen,Garbrecht:2025crs,Bao:2026bgu}. }  
Fluctuations of the spectator modify the bounce action, on which the nucleation rate depends exponentially. The resulting spatial variations in the timing of vacuum-energy release, which inherit a nearly scale-invariant spectrum, generate large-scale curvature perturbations. The modulation also changes the bubble population and characteristic scale, directly affecting the GW source. The resulting intrinsic GW anisotropies reflect the primordial isocurvature fluctuations and provide a signal correlated with the curvature perturbations measured in the CMB.\footnote{GWs sourced at second order by curvaton isocurvature perturbations have also been studied~\cite{Bartolo:2007vp}; here the GW source is the modulated FOPT.}

We illustrate this mechanism through a supercooled FOPT~\cite{Guth:1980zk,Witten:1980ez}. We consider a classically scale-invariant (CSI) sector, in which radiative symmetry breaking generates the transition scale~\cite{Coleman:1973jx,Gildener:1976ih}, and introduce a spectator-dependent fermion coupling that modulates the bounce action while the spectator remains effectively frozen. This realization reproduces the observed amplitude of large-scale curvature perturbations and produces a loud stochastic GW background in the millihertz band. However, the intrinsic GW anisotropies are still modest. We then introduce a positive scalar mass term that changes the nucleation history and enhances the GW source response near the boundary for transition completion. We identify a benchmark with an intrinsic fractional GW modulation of roughly thirty times the curvature amplitude. For this benchmark, our forecast shows a quadrupole signal-to-noise ratio (SNR) of order unity for a five-year LISA--Taiji observation.
The predicted local primordial non-Gaussianity is within reach of next-generation large-scale-structure surveys~\cite{Euclid:2025hlc,SPHEREx:2014bgr}.

\textbf{\textit{Curvature and GW response from a modulated FOPT.}}---
On superhorizon scales, each patch evolves as a separate homogeneous universe labeled by $\xi=\sigma/\Lambda$, 
where $\sigma$ is the value of the spectator field and $\Lambda$ is a scale set by the coupling between the spectator field and the PT field. We assume a frozen, energetically negligible spectator and initially unperturbed radiation. Define the local e-fold number $N=\ln(a/a_{\mathrm i})$, where $a$ is the scale factor and $a_{\mathrm i}$ its value on a common initial flat slice before nucleation. The e-fold number $N_{\mathrm f}(\xi)$ to a final slice of uniform density $\rho_{\mathrm f}$ after conversion to radiation determines the curvature perturbation $\zeta$ via the $\delta N$ formalism~\cite{Sasaki:1995aw,Lyth:2004gb}:
\begin{equation}
 \zeta(\boldsymbol{x})=\delta N_{\mathrm f}
 =\mathcal N_1\delta\xi
 +\frac12\mathcal N_2\delta\xi^2
 +\cdots , \label{eq:deltaN-map}
\end{equation}
where $\delta N_{\mathrm f}\equiv N_{\mathrm f}(\xi)-N_{\mathrm f}(\bar\xi)$, $\mathcal N_n\equiv\left.(\D ^nN_{\mathrm f}/\D\xi^n)\right|_{\xi = \bar\xi}$, and $\delta\xi(\boldsymbol{x})\equiv\xi(\boldsymbol{x})-\bar\xi$. Here $\boldsymbol{x}$ denotes comoving position and $\bar\xi\equiv \bar{\sigma}/\Lambda$ the homogeneous background value. All derivatives are evaluated at $\bar\xi$, holding other model parameters and the initial slice fixed.

We compute $\zeta$ from the PT history, described by the false-vacuum fraction $P_{\mathrm F}(N,\xi)=\e^{-I(N,\xi)}$~\cite{Guth:1982pn}, where $I$ is the bubble nucleation-and-growth exponent (see Appendix~\ref{app:bubble-history}). For monotonic conversion, the converted fraction $u=1-P_{\mathrm F}$ labels hypersurfaces $N=N_u(\xi)$ satisfying
\begin{equation}
 I(N_u,\xi)=-\ln(1-u),\quad
 \partial_\xi N_u=-\left.\frac{\partial_\xi I}{\partial_N I}\right|_{N=N_u} ,
 \label{eq:transition-surface-response}
\end{equation}
where $\partial_\xi N_u$ keeps $u$ fixed, while partial derivatives of $I$ keep its other argument fixed.

For a constant, $\xi$-independent  vacuum-energy difference $\rho_V$, we assume that the released energy density $\rho_{\rm rel}$ redshifts as radiation.\footnote{Slower reheating may allow an early matter-dominated era with distinct cosmological consequences~\cite{Ai:2026zrs}, although its onset depends on the post-collision scalar field dynamics~\cite{Mansour:2026sdx}.} The pre-existing radiation evolves as $\rho_r=\rho_{r,\mathrm i}\e^{-4N}$ with initial density $\rho_{r,\mathrm i}$. As we assume a negligible initial perturbation in the radiation, $\rho_{r,\rm i}$ is also homogeneous. Energy conservation implies
\begin{equation}
 \rho_{\rm rel}'+4\rho_{\rm rel}=-\rho_VP_{\mathrm F}',
 \label{eq:continuous-conversion}
\end{equation}
where a prime denotes $\D/\D N$, with initial values $P_{\mathrm F}=1$ and $\rho_{\rm rel}=0$. After completion, $\rho_{\mathrm f}=\rho_r(N_{\mathrm f})+\rho_{\rm rel}(N_{\mathrm f})$. Integrating Eq.~\eqref{eq:continuous-conversion} yields
\begin{equation}
 \mathcal E(\xi)\equiv\rho_{\mathrm f} \e^{4N_{\mathrm f}(\xi)}
 =\rho_{r,\mathrm i}+\rho_V\int_0^1 \D u\,\e^{4N_u(\xi)}.
 \label{eq:release-functional}
\end{equation}
A later release corresponds to larger $N_u$ at fixed $u$ and hence larger $N_\mathrm{f}(\xi)$. Physically, since the vacuum energy density does not redshift, delaying its conversion to radiation requires more expansion to reach the final density $\rho_\mathrm{f}$. Let $\bar{\mathcal E}=\mathcal E(\bar\xi)$, $\bar N_u=N_u(\bar\xi)$. The resulting curvature perturbation is
\begin{equation}
 \zeta=\frac{1}{4}\ln \frac{\mathcal{E}}{\bar{\mathcal{E}}}=\frac14\delta\ln\mathcal E.
 \label{eq:continuous-curvature}
\end{equation}
The linear and quadratic responses of $\zeta$ to the modulation are, respectively,
\begin{subequations}
\label{eq:curvature-responses}
\begin{align}
 \mathcal N_1 &=\frac{\rho_V}{\bar{\mathcal E}}
 \int_0^1\D u\,\e^{4\bar N_u}\partial_\xi N_u,\\
 \mathcal N_2 &=\frac{\rho_V}{\bar{\mathcal E}}
 \int_0^1\D u\,\e^{4\bar N_u}
 \left[\partial_\xi^2N_u+4(\partial_\xi N_u)^2\right]-4\mathcal N_1^2.
\end{align}
\end{subequations}

In the sudden-transition limit, all vacuum energy is released on a single hypersurface $N = N_{\rm PT}(\xi)$. The integral in Eq.~\eqref{eq:release-functional} then reduces to $\e^{4N_{\rm PT}(\xi)}$ and we get
\begin{equation}
 \zeta=\frac14\ln\frac{1+\bar\alpha\, \e^{4\tau}}{1+\bar\alpha}
 =\frac{\bar\alpha}{1+\bar\alpha}\tau+O(\tau^2),
 \label{eq:sudden-conversion-picture}
\end{equation}
where $\tau\equiv \delta N_{\rm PT}(\xi)$ is the hypersurface displacement and $\bar\alpha\equiv\rho_V/\rho_r(\bar N_{\rm PT})$ is the PT strength.
The conversion is inefficient for $\bar\alpha\ll1$, while $\zeta\simeq\tau$ 
for sufficiently strong PTs~\cite{Bartolo:2003bz,Bartolo:2003jx,Lyth:2005qk}. For a radiation bath with temperature $T\propto a^{-1}$, $\tau=\ln[\bar T_{\rm PT}/T_{\rm PT}(\xi)]\simeq-\delta T_{\rm PT}/\bar T_{\rm PT}$ at linear order, where $T_{\rm PT}(\xi)$ is the radiation temperature just before release and $\delta T_{\rm PT}\equiv T_{\rm PT}(\xi)-\bar T_{\rm PT}$. For a generic PT of finite duration, one can obtain $\zeta$ by numerically evaluating the history integral in Eq.~\eqref{eq:release-functional}.

Assuming Gaussian fluctuations of $\xi$, the dimensionless power spectrum is $P_{\delta\xi}(k)=[H_{\rm inf}(k)/(2\pi\Lambda)]^2$, where $k$ is the comoving wavenumber and $H_{\rm inf}(k)$ the inflationary Hubble rate at horizon exit. If $\xi$ accounts for all observed curvature perturbations, the tree-level power and local bispectrum amplitude are~\cite{Lyth:2005fi}
\begin{equation}
 P_\zeta=\mathcal N_1^2P_{\delta\xi},\quad
 f_{\rm NL}^{\rm loc}=\frac56\frac{\mathcal N_2}{\mathcal N_1^2}.
 \label{eq:scalar-statistics}
\end{equation}
The origin of non-Gaussianity is readily seen in the sudden-transition limit, where Eq.~\eqref{eq:sudden-conversion-picture} gives
\begin{equation}
 f_{\rm NL}^{\rm loc}
 =\frac{10}{3\bar\alpha}
 +\frac{5(1+\bar\alpha)}{6\bar\alpha}
 \left.\frac{\partial_\xi^2N_{\rm PT}}{(\partial_\xi N_{\rm PT})^2}\right|_{\xi=\bar\xi}.
 \label{eq:sudden-fnl-decomposition}
\end{equation}
The first term reflects nonlinear entropy-to-curvature conversion, analogous to that in the conventional curvaton mechanism~\cite{Lyth:2005fi}, while the second term arises from nonlinear modulation of the transition hypersurface.

The GW energy-density spectrum emitted during the FOPT is expressed as $\Omega_{\rm GW}(f;\xi)=\rho_{c,0}^{-1}\D \rho_{\rm GW,0}/\D \ln f$, with $\rho_{c,0}$ being the common present critical density, $\rho_{\rm GW,0}$ the GW energy density today, and $f$ the observed frequency. The unperturbed part is $\bar\Omega_{\rm GW}(f)\equiv{\Omega_{\rm GW}(f;\bar\xi)}$. The total fractional anisotropy of the observed GW energy-density spectrum reads
\begin{equation}
 \begin{aligned}
 \delta_{\rm GW}^{\rm tot}(f,\hat{\boldsymbol n})
 &\equiv
 \frac{\Omega_{\rm GW}^{\rm obs}(f,\hat{\boldsymbol n})
 -\bar\Omega_{\rm GW}(f)}{\bar\Omega_{\rm GW}(f)},
 \end{aligned}
 \label{eq:total-gw-anisotropy}
\end{equation}
where $\hat{\boldsymbol n}$ is the sky direction. The directional spectrum is normalized so that its unperturbed isotropic limit is $\bar\Omega_{\rm GW}(f)$. At linear order, the total anisotropy can be decomposed into the intrinsic source and propagation contributions $\delta_{\rm GW}^{\rm tot}=\delta_{\rm GW}^{\rm src}(f,\boldsymbol x_{\mathrm f})+\delta_{\rm GW}^{\rm prop}(f,\hat{\boldsymbol n})$, where $\boldsymbol x_{\mathrm f}$ is the intersection of the unperturbed line of sight with a final uniform-density slice after GW production has ended. The intrinsic source contribution can be obtained by evaluating the GW spectrum on this final slice and applying a common background redshift from that slice to today. To linear order,
\begin{equation}
 \begin{aligned}
 \delta_{\rm GW}^{\rm src}(f,\boldsymbol x_{\mathrm f})
 =\left.\frac{\partial\ln{\Omega_{\rm GW}}}{\partial\xi}
 \right|_{\xi=\bar\xi}
 \delta\xi(\boldsymbol x_{\mathrm f})\equiv\mathcal T_{\rm src}(f)\zeta(\boldsymbol x_{\mathrm f}).
 \end{aligned}
 \label{eq:common-source-response}
\end{equation}
The same $\delta\xi$ thus sources the intrinsic GW anisotropy and the curvature perturbation that seeds the CMB anisotropies. For a fixed spectral shape $S$, $\Omega_{\rm GW}(f;\xi)=\Omega_{\rm peak}(\xi)S[f/f_{\rm peak}(\xi)]$, with peak amplitude $\Omega_{\rm peak}$, peak frequency $f_{\rm peak}$, and $S(1)=1$. Then
\begin{align}
\label{eq:spectral-response}
    \mathcal T_{\rm src}(f) =\frac{G_\Omega-n_\Omega(f)G_f}{\mathcal N_1},
\end{align} 
where $G_\Omega\equiv{\left.\partial_\xi\ln\Omega_{\rm peak}\right|_{\xi=\bar\xi}}$, $n_\Omega\equiv\partial\ln\bar\Omega_{\rm GW}/\partial\ln f$, and $G_f\equiv{\left.\partial_\xi\ln f_{\rm peak}\right|_{\xi=\bar\xi}}$. Curvature responds to the history of energy release, whereas GWs also respond to the PT dynamics. The coefficient $\mathcal T_{\rm src}$ measures the intrinsic fractional GW response relative to $\zeta$.

The propagation contribution is~\cite{Alba:2015cms,Contaldi:2016koz,Bartolo:2019oiq,Bartolo:2019zvb,Malhotra:2022ply}
\begin{equation}
 \delta_{\rm GW}^{\rm prop}(f,\hat{\boldsymbol n})
 \equiv\frac{\Omega_{\rm GW}^{\rm obs}(f,\hat{\boldsymbol n})
 -\Omega_{\rm GW}[f;\xi(\boldsymbol x_{\mathrm f})]}
 {\bar\Omega_{\rm GW}(f)}.
 \label{eq:propagation-anisotropy}
\end{equation}
We focus on the large angular scales probed by GW detectors, corresponding to adiabatic growing modes that remain superhorizon through radiation--matter equality. For these
modes, and neglecting anisotropic stress, late-time integrated Sachs--Wolfe (ISW) and Doppler effects, the propagation
contribution reduces to the standard matter-era Sachs--Wolfe (SW) plateau, $-(4-n_\Omega)\zeta/5$, so that the total anisotropy is approximately
\begin{equation}
 \delta_{\rm GW}^{\rm tot}(f,\hat{\boldsymbol n})
 \simeq\left[\mathcal T_{\rm src}(f)-\frac{4-n_\Omega(f)}{5}\right]
 \zeta(\boldsymbol x_{\mathrm f}).
 \label{eq:source-plus-sw}
\end{equation}
For spectral slopes of order unity, the combined propagation contribution produces a fractional anisotropy of order $\zeta$, whereas the modulated PT can yield $|\mathcal T_{\rm src}|\gg1$ and substantially enhance it. See Appendix~\ref{app:gw-response} for details.

\textbf{\textit{A scale-invariant realization.}}---
A supercooled PT is particularly well suited to realizing our mechanism for three reasons: i) vacuum domination makes the conversion efficient, ii) cooling suppresses thermal corrections to the spectator mass, and iii) it can enhance both the GW monopole and its anisotropies. Specifically, we consider a complex scalar $\Phi$ charged under $U(1)_D$, with gauge coupling $g_D$ and radial field $\phi=\sqrt{2}|\Phi|$. With no tree-level quadratic term, radiative symmetry breaking generates the vacuum expectation value $v_\phi$~\cite{Hempfling:1996ht,Chang:2007ki,Gouttenoire:2023pxh}. The gauge contribution to the Coleman--Weinberg potential is~\cite{Coleman:1973jx,Khoze:2013uia}
\begin{equation}
 V_{\rm CW}^{(g)}(\phi)=\frac{\beta_\lambda}{4}\phi^4
 \left[\ln\left(\frac{\phi}{v_\phi}\right)-\frac14\right]
 +\frac{\beta_\lambda v_\phi^4}{16},
 \label{eq:csi-cw-potential}
\end{equation}
where $\beta_\lambda\simeq6(g_D^2/4\pi)^2$. Eq.~\eqref{eq:csi-cw-potential} implies a zero-energy vacuum at
$\phi=v_\phi$ and an energy density
$\rho_V=\beta_\lambda v_\phi^4/16$ at the symmetric point $\phi=0$.

We introduce a Dirac fermion $\psi$ with an effective coupling $-m_\psi(\phi,\xi)\bar\psi\psi$. For simplicity, we take $m_\psi^2(\phi,\xi)=\xi^2 v_\phi^2 F(\phi)$. For a fixed value of $\phi$, this term can induce a mass for the spectator field $\sigma$ via the fermion loop. To ensure that the $\sigma$ field is effectively frozen before the PT and that it does not have a large potential energy after the PT, we impose $F(0)=F(v_\phi)=\partial_\phi F(v_\phi)=0$. As an example, we take $F(\phi)=(\phi^2/v_\phi^2)(1-\phi^2/v_\phi^2)^2$.
Although the fermion one-loop corrections induce no spectator mass at $\phi=0$, thermal fluctuations of $\phi$ generate a thermal mass $m_{\sigma,\rm th}^2\simeq T^4/(72\Lambda^2)$ and drive the spectator toward the origin (see Appendix~\ref{app:csi-consistency}). To keep the spectator effectively frozen, we require
\begin{equation}
 |m_{\sigma,\rm eff}^2|\ll H^2,\quad
 |\Delta\sigma| \ll |\bar\sigma|,
 \label{eq:csi-spectator-conditions}
\end{equation}
until the conversion is complete, where $m_{\sigma,\rm eff}$ is the effective mass of $\sigma$ and $\Delta\sigma$ the accumulated displacement.

Our calculation distinguishes two time scales. We resolve the nucleation-and-growth history to determine the relevant PT parameters, including the transition strength and mean bubble separation. On longer cosmological time scales, we adopt the sudden-transition approximation, using these parameters as inputs to compute the resulting curvature perturbations and GW responses. We use \texttt{CosmoTransitions}~\cite{Wainwright:2011kj} to numerically solve the $O(3)$ bounce, whose three-dimensional Euclidean action $S_3$ determines the thermal tunneling exponent $B\equiv S_3/T$. We then integrate the bubble moments and number density as described in Appendix~\ref{app:bubble-history}. 
Percolation is defined by
$I(N_p,\xi)=0.34$, and we identify $N_{\rm PT}=N_p$.
The curvature response follows from Eq.~\eqref{eq:sudden-conversion-picture}.

We require the PT to reach $I=4.6$, i.e.~a false-vacuum fraction of about $1\%$, as our completion criterion. Completion also requires the physical false-vacuum volume to shrink, which occurs for $q\equiv\partial_N I>3$~\cite{Ellis:2018mja,Athron:2022mmm}. To favor the applicability of the sound-wave template, we impose the more conservative cuts
$q_{p}>4.5$ and $H_{p}R_{*}<0.4$, where $q_p\equiv q(N_p)$ is the growth rate of the exponent
$I$, and $H_{p}$ and $R_*$ are the Hubble rate and mean bubble separation, all evaluated at
percolation, $I(N_p)=0.34$. Specifically, $R_{*}=n_b(N_{p})^{-1/3}$, where $n_b$ is the physical bubble number density. 
Varying $\xi$, we extract $\mathcal N_{1,2}$ from Eq.~\eqref{eq:sudden-conversion-picture} and the GW response from the acoustic template in Appendix~\ref{app:gw-response}.

The model has three main parameters, $g_D$, $v_\phi$, and $\bar\xi$, while the observed curvature power $P_\zeta=2.1\times10^{-9}$ fixes $H_{\rm inf}/\Lambda$. We take $H_{\rm inf}=2\times10^{13}\,\mathrm{GeV}$ throughout. With tensor power $P_T=2H_{\rm inf}^2/(\pi^2m_{\rm Pl}^2)$, this gives a tensor-to-scalar ratio $r\equiv P_T/P_\zeta\simeq6.5\times10^{-3}$~\cite{Planck:2018jri}. This value satisfies the bound $r_{0.05}<0.036$ at $95\%$ confidence, evaluated at the pivot scale $k=0.05\,\mathrm{Mpc}^{-1}$~\cite{Planck:2018jri,BICEP:2021xfz}, and lies within the forecast sensitivity of next-generation CMB experiments~\cite{LiteBIRD:2022cnt,CMB-S4:2020lpa}. During vacuum domination, Eq.~\eqref{eq:acoustic-peak-frequency} gives $f_{\rm peak}\propto g_Dv_\phi/(H_{p}R_{*})$. Thus, at fixed dimensionless parameters, $v_\phi$ sets the frequency scale, with $v_\phi=\mathcal O(10\,\mathrm{TeV})$ placing the peak in the millihertz band accessible to LISA~\cite{LISA:2017pwj}. Taking a bubble-wall speed $v_w\approx 1$ and effective relativistic degrees of freedom $g_{\star\rho}=g_{\star s}=100$ for the energy and entropy densities, respectively, we consider the benchmark
\begin{equation}
 g_D=0.745,\quad v_\phi=10\,\mathrm{TeV},\quad
 \bar\xi=0.536 ,
 \label{eq:csi-benchmark}
\end{equation}
which yields a critical temperature $T_c=2.38\,\mathrm{TeV}$ and a percolation temperature $T_{p}=2.36\,\mathrm{GeV}$, with $q_p=4.52$ and $H_{p}R_{*}=0.320$. The curvature response and local non-Gaussianity are $\mathcal N_1=47.1$ and $f_{\rm NL}^{\rm loc}=0.277$. Matching the observed $P_\zeta$ fixes $\Lambda=3.27\times10^{18}\,\mathrm{GeV}$, for which the spectator thermal mass satisfies $m_{\sigma,\rm th}^2/H^2\simeq7.0\times10^{-4}$ at $T_c$. The $O(4)$ quantum-tunneling action $S_4\simeq705$ is much larger than the thermal tunneling exponent $B=S_3/T_{p}\simeq114$ at percolation, supporting the dominance of thermal nucleation.

\begin{figure}[t!]
\centering
\includegraphics[width=\columnwidth]{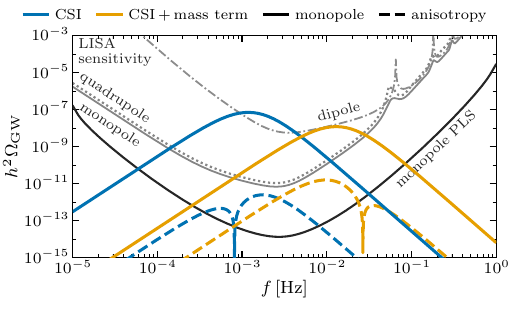}
\caption{GW monopole (solid) and absolute quadrupole anisotropy (dashed) for the CSI benchmark (blue) and the benchmark with the mass term (orange). Gray: LISA instrument-noise sensitivity to the monopole, dipole and quadrupole~\cite{LISACosmologyWorkingGroup:2022kbp}; black: monopole power-law integrated sensitivity (PLS)~\cite{Thrane:2013oya} ($T_{\rm obs}=3\,$yr, ${\rm SNR}=1$). }
\label{fig:sensitivity}
\end{figure}

The GW monopole, corresponding to the sky-averaged spectrum, is readily detectable. The acoustic template in Appendix~\ref{app:gw-response} gives a peak amplitude $h^2\Omega_{\rm peak}=7.0\times10^{-8}$ at $f_{\rm peak}\simeq1.19\,\mathrm{mHz}$. The peak lies nearly four orders of magnitude above the adopted LISA noise curve (see Fig.~\ref{fig:sensitivity}).
For an observation time $T_{\rm obs}=3\,\mathrm{yr}$, the monopole reaches ${\rm SNR}_0\simeq1.3\times10^3$ (see Appendix~\ref{app:gw-response} for details). The measurement is limited primarily by statistical fluctuations of the GW background itself rather than instrumental noise.

To assess the detectability of GW anisotropies, we use this benchmark's monopole spectrum and the angular power predicted by Eq.~\eqref{eq:angular-power}. We calculate the SNRs using the detector response described in Ref.~\cite{LISACosmologyWorkingGroup:2022kbp}, including these GW background fluctuations in the full detector covariance~\cite{Mentasti:2023icu,Mentasti:2023uyi}.
The quoted angular-mode SNRs assume that the GW sky pattern is known, with its amplitude inferred from GW data. Because the GW anisotropy and the CMB share the primordial source $\zeta$, cross-correlating with the CMB may help constrain this pattern and improve sensitivity over a GW angular auto-power search.
The template accuracy depends on the angular correlation after propagation and projection.

For the CSI benchmark, the total response $\mathcal T_{\rm tot}$ remains of order unity in the relevant frequency band. The characteristic fractional anisotropy is of order $2\times10^{-5}$. Over the same observing period, the resulting angular-mode SNRs are ${\rm SNR}_{\ell=2}\simeq 1.9\times10^{-2}$ for the quadrupole and ${\rm SNR}_{\ell=1}\simeq 1.6\times10^{-3}$ for the dipole. Detector networks can improve the sensitivity to GW anisotropies~\cite{Baker:2019ync,Cui:2023dlo,Mentasti:2023uyi,Zhao:2024yau,Li:2025pde} (see Ref.~\cite{Zhou:2023rop} for TianQin~\cite{TianQin:2015yph}). Adding Taiji~\cite{Hu:2017mde} and including the LISA--Taiji signal correlations raises these SNRs to $2.9\times10^{-2}$ and $8.4\times10^{-3}$, respectively. Both remain well below unity.
Keeping the monopole spectrum and the frequency shape of $\mathcal T_{\rm tot}$ fixed, reaching ${\rm SNR}=1$ for the quadrupole would require increasing the normalization of $\mathcal T_{\rm tot}$ by factors of approximately $53$ for LISA and $34$ for the LISA--Taiji network, relative to this benchmark.

The size of these statistical fluctuations can be estimated by counting the independent samples of the stochastic background. For an effective observing bandwidth $\Delta f\sim10^{-2}\,\mathrm{Hz}$, their number is approximately $N_{\rm modes}\sim2T_{\rm obs}\Delta f\sim2\times10^6$. This gives a characteristic fractional sampling uncertainty of order $N_{\rm modes}^{-1/2}\sim7\times10^{-4}$, larger than the fractional anisotropy predicted for this benchmark. Note that this mode-counting argument is only an estimate, as the precise multipole sensitivity depends on the detector response and the full channel covariance. For a fixed observing configuration and bandwidth, reducing instrumental noise alone cannot eliminate this statistical uncertainty.

Thus, although this benchmark reproduces the observed curvature power and yields a detectable GW monopole, its GW anisotropies are not detectable under the adopted observing assumptions.
Physically, the modulation mainly shifts the end of the long supercooling stage, characterized by $\ln(T_c/T_{p})\simeq6.91$, while producing a much smaller relative response in the dimensionless bubble scale, $\partial_\xi\ln(H_{p}R_{*})/\mathcal N_1\simeq0.20$. Meanwhile, the efficiency of converting released vacuum energy into bulk fluid motion is nearly saturated, and the post-transition radiation temperature is determined predominantly by the $\xi$-independent $\rho_V$. The large timing response therefore generates the observed curvature perturbations from small $\delta\xi$, without a comparably large response of the GW source.
Motivated by this limitation, we modify the minimal CSI model below to enhance its intrinsic GW anisotropy.

\textbf{\textit{Enhancement of GW response near the completion boundary.}}---We add a positive mass term $+M^2\phi^2/2$ to the potential while preserving zero vacuum energy at $v_\phi$,
\begin{equation}
 V_{\rm CW}^{(g)}\rightarrow V_{\rm CW}^{(g)}
 -\frac{M^2}{4v_\phi^2}(\phi^2-v_\phi^2)^2.
 \label{eq:positive-mass-deformation}
\end{equation}
The vacuum-energy difference remains independent of $\xi$ and is given by $\rho_V=\beta_\lambda v_\phi^4(1-4r_M)/16$, where $r_M\equiv M^2/(\beta_\lambda v_\phi^2)>0$.
The bounce action $B(T)$ then develops a minimum, so further cooling eventually suppresses thermal nucleation~\cite{Levi:2022bzt}. When the model parameters lie close to the boundary for successful transition completion, a small modulation of $B(T)$ can appreciably change the accumulated bubble number and the growth required for percolation. The resulting response of $H_{p}R_{*}$ affects both the GW amplitude and peak frequency and can therefore enhance the intrinsic GW anisotropy.

After computing the full history near the completion boundary specified above, we find the benchmark
\begin{equation}
 g_D=1.3,\ v_\phi=60\,\mathrm{TeV},\
 \bar\xi=0.85,\ r_M=0.025,
 \label{eq:enhanced-benchmark}
\end{equation}
corresponding to $M=3.13\,\mathrm{TeV}$, $T_{p}=6.68\,\mathrm{TeV}$, $q_{p}=5.71$, $H_{p}R_{*}=0.351$ and $\mathcal N_1=5.34$. Again, $\Lambda=3.71\times10^{17}\,\mathrm{GeV}$ is fixed by the observed $P_\zeta$. The local bispectrum amplitude is $f_{\rm NL}^{\rm loc}=9.48$. The constraint from Planck indicates $f_{\rm NL}^{\rm loc}=-0.9\pm5.1$ at $68\%$ confidence~\cite{Planck:2019kim}. Hence this benchmark stands near the $95\%$ confidence boundary and within reach of next-generation large-scale-structure surveys~\cite{Euclid:2025hlc,SPHEREx:2014bgr}.  Furthermore, the leading thermal correction is $m_{\sigma,\rm th}^2/H^2=0.054$ at $T_c=25.5\,\mathrm{TeV}$. Evolving the spectator under this correction from $T_c$ with $\sigma'=0$ leads to a drift of $2.8\%$ before percolation. We find $S_4\simeq891$ from the $O(4)$ calculation, well above $B\simeq128$ at percolation. Interestingly, we arrive at $\partial_\xi\ln(H_{p}R_{*})/\mathcal N_1=9.15$ for the bubble-scale response, much larger than the CSI case. 

Using the acoustic template in the sudden-transition approximation (Appendix~\ref{app:gw-response}), we obtain the fraction of the total energy in bulk fluid motion, $K_*=0.326$, and the suppression factor associated with the finite lifetime of the sound-wave source $\Upsilon_{\rm sw}=0.357$. The corresponding peak amplitude and frequency are $h^2\Omega_{\rm peak}=1.20\times10^{-8}$ and $f_{\rm peak}=12.9\,\mathrm{mHz}$. The responses are $G_\Omega/\mathcal N_1=24.58$ and $G_f/\mathcal N_1=-9.98$, so $\mathcal T_{\rm src}(f)\simeq24.58+9.98\, n_\Omega(f)$. The normalized peak-amplitude response $G_\Omega/\mathcal N_1$ receives contributions $9.15$, $12.21$, and $3.22$ from $H_{p}R_{*}$, $K_*^2$, and $\Upsilon_{\rm sw}$, respectively, indicating that both the bubble scale and fluid efficiency contribute substantially to the enhanced GW response.

The intrinsic absolute fluctuation, $h^2\bar\Omega_{\rm GW}|\mathcal T_{\rm src}|\sqrt{P_\zeta}$, is largest at $f\simeq 9.81\,\mathrm{mHz}$, where it reaches $1.63\times10^{-11}$. At this frequency, Eq.~\eqref{eq:source-plus-sw} gives
\begin{equation}
 \delta_{\rm GW}^{\rm tot}\simeq(33.4-0.6)\zeta
 =32.8\zeta.
 \label{eq:benchmark-signed-anisotropy}
\end{equation}
The propagation term $-(4-n_\Omega)/5\simeq-0.6$ partially cancels the intrinsic source, leaving an enhancement of $\sim30$ relative to $\zeta$. The coherent absolute fluctuation peaks at $9.74\,\mathrm{mHz}$, reaching $1.60\times10^{-11}$. A $5\,$yr observation with the LISA--Taiji network reaches $\mathrm{SNR}_{\ell=2}\simeq 1.05$ assuming the sky pattern is known. As summarized in Fig.~\ref{fig:sensitivity}, despite having a monopole peak about six times lower, this benchmark has a peak absolute quadrupole amplitude about six times higher than that of the CSI benchmark.

\textbf{\textit{Discussion and conclusions.}}---In this Letter, we have proposed a new mechanism for generating large-scale curvature perturbations through a {\it modulated FOPT} that occurs after inflation. The order-parameter field couples to a light spectator, which acquires fluctuations during inflation and converts them into adiabatic curvature perturbations by modulating the bubble nucleation rate during the PT. We have illustrated this mechanism in a simple CSI model, which also produces a GW monopole detectable by future space-based experiments such as LISA. An interesting feature is the accompanying GW anisotropy, which is correlated with the primordial curvature perturbations that seed the CMB anisotropies. Although this anisotropy is too small to be detectable for our minimal CSI benchmark, adding a positive scalar mass can substantially enhance the GW response. We identify a benchmark whose anisotropy reaches ${\rm SNR}_{\ell=2}\simeq1$ under the adopted forecast assumptions and whose local primordial non-Gaussianity is within reach of future large-scale-structure surveys. These initial results motivate further exploration of modulated FOPTs as an origin of primordial perturbations with correlated GW signatures.

\textbf{\textit{Acknowledgments.}}---
We acknowledge the use of Claude (Anthropic) and Codex (OpenAI) to help with writing and running code, exporting figures, and draft proofreading.
The authors directed, checked, and revised all outputs included in this work.
G.F. acknowledges support from the Italian Ministero dell'Università e della Ricerca through the FIS 3 project FIS-2024-02546 “PRIMAVERA” (CUP C53C25001070001), the Departments of Excellence grant 2023-2027 “Quantum Frontier”, as well as from Istituto Nazionale di Fisica Nucleare (INFN) through the Theoretical Astroparticle Physics (TAsP) project. X.W. is funded by the European Union, NextGenerationEU, National Recovery and Resilience Plan (mission 4, component 2)
under the project {\it MODIPAC: Modular Invariance in Particle Physics and Cosmology} (CUP C93C24004940006).

\clearpage
\appendix
\setcounter{secnumdepth}{1}

\section{Bubble nucleation and growth}
\label{app:bubble-history}

\textbf{Background and nucleation.} To determine the curvature and GW responses, we first evolve the bubble population and energy release in each patch. The continuous background obeys
\begin{equation}
 \rho_{\rm tot}=\rho_r+\rho_VP_{\mathrm F}+\rho_{\rm rel},\quad
 H^2=\frac{\rho_{\rm tot}}{3m_{\rm Pl}^2},
 \label{eq:continuous-friedmann}
\end{equation}
where $H$ is the Hubble rate and $m_{\rm Pl}$ the reduced Planck mass. We take $T=T_{\mathrm i}\e^{-N}$, where $T_{\mathrm i}$ is the initial temperature, neglecting reheating of the bath during the transition. The continuous background calculation includes the contribution of the released energy to $H$. On longer cosmological time scales, we use the sudden-transition approximation, where the continuous energy release is neglected, and the nucleation and bubble-growth history is evaluated on the unheated background $H^2=(\rho_r+\rho_V)/(3m_{\rm Pl}^2)$.
The thermal nucleation rate per unit physical false-vacuum volume and proper time is
\begin{equation}
 \Gamma=T^4\left(\frac{B}{2\pi}\right)^{3/2}\e^{-B},
 \quad B\equiv\frac{S_3}{T},
 \label{eq:thermal-rate-convention}
\end{equation}
where $S_3$ is the three-dimensional Euclidean bounce action~\cite{Linde:1981zj}. 

For quantum tunneling at low temperatures, it is also necessary to check the $O(4)$-symmetric Euclidean bounce action $S_4$~\cite{Coleman:1977py,Callan:1977pt} and estimate the corresponding rate as
\begin{equation}
 \Gamma_4\simeq R_4^{-4}\left(\frac{S_4}{2\pi}\right)^2\e^{-S_4},
 \label{eq:quantum-rate-diagnostic}
\end{equation}
where $R_4$ is the half-maximum radius of the $O(4)$ bounce profile. In our analysis, we compare Eq.~\eqref{eq:quantum-rate-diagnostic} with the thermal rate in Eq.~\eqref{eq:thermal-rate-convention} to check whether thermal nucleation dominates.

\textbf{Bubble growth.}
Once a bubble nucleates at proper time $t_n$, its physical radius at a later time $t$ is~\cite{Guth:1982pn}
\begin{equation}
 R(t,t_n)=a(t)\int_{t_n}^t\frac{v_w(\widetilde t)}{a(\widetilde t)}\D \widetilde t,
 \label{eq:bubble-growth-general}
\end{equation}
where $v_w(t)$ is the physical wall speed and the initial bubble radius has been neglected. To evolve the full nucleation history, we define its physical-radius moments
\begin{equation}
 M_j(t,\xi)=\int_{t_{\mathrm i}}^t \D t_n\,\Gamma(t_n,\xi)
 \left[\frac{a(t_n)}{a(t)}\right]^3R^j(t,t_n),
 \label{eq:bubble-moment-definition}
\end{equation}
where $j=0,\ldots,3$ and $t_{\mathrm i}$ is the initial proper time. Using $\D R/\D t=HR+v_w$, we obtain the following equations for $M_j$:
\begin{subequations}
\label{eq:bubble-moment-odes}
\begin{align}
 M_0'+3M_0&=\frac{\Gamma}{H},\\
 M_j'+(3-j)M_j&=j\frac{v_w}{H}M_{j-1},\quad j=1,2,3.
\end{align}
\end{subequations}
The conversion exponent and its growth rate follow as
\begin{equation}
 I=\frac{4\pi}{3}M_3,\quad q\equiv I'=4\pi\frac{v_w}{H}M_2.
 \label{eq:bubble-functional}
\end{equation}
Note that these moments contain no $P_{\mathrm F}$ factor. The physical bubble number density instead obeys~\cite{Athron:2022mmm}
\begin{equation}
 n_b'+3n_b=\frac{\Gamma P_{\mathrm F}}{H},\quad {R_*=n_b(N_p)^{-1/3}},
 \label{eq:bubble-number}
\end{equation}
where $R_*$ is the mean bubble separation at percolation.
Integrating the above moment and number-density equations
on the appropriate background specified above gives the
full nucleation and bubble-growth history in each patch. By comparing these histories at different $\xi$, we then determine the curvature and GW responses to the modulation.

\section{GW spectrum and anisotropies}
\label{app:gw-response}

\textbf{Acoustic spectrum.} For the sudden benchmarks, we estimate the acoustic spectrum using the bubble separation at percolation~\cite{Hindmarsh:2015qta,Caprini:2015zlo,Caprini:2019egz}. Define the strength $\alpha_\theta=[\Delta F-(T/4)\partial_{T}\Delta F]/\rho_r$, 
where $\Delta F\equiv F_{\rm false}-F_{\rm broken}$ is the free-energy density difference between the two phases and the derivative holds $\xi$ fixed. All source quantities below are evaluated at percolation. The fraction of total energy in fluid motion is approximated by~\cite{Espinosa:2010hh}
\begin{equation}
 K_*=\frac{\kappa_{\rm sw}\alpha_\theta}{1+\alpha_\theta},\quad
 \kappa_{\rm sw}=\frac{\alpha_\theta}{0.73+0.083\sqrt{\alpha_\theta}+\alpha_\theta},
 \label{eq:acoustic-efficiency}
\end{equation}
with $\kappa_{\rm sw}$ the conversion efficiency into bulk motion. We take the proper acoustic lifetime $\tau_{\rm sw}=R_{*}/\bar U_f$, where we estimate the root-mean-square fluid velocity using $\bar U_f^2\simeq 3K_*/4$. The finite-duration factor accounting for radiation-era expansion is~\cite{Ellis:2020awk,Guo:2020grp}
\begin{equation}
 \Upsilon_{\rm sw}=1-\frac{1}{\sqrt{1+2H_{p}\tau_{\rm sw}}}.
 \label{eq:acoustic-lifetime}
\end{equation}
The peak amplitude and frequency are
\begin{align}
 h^2\Omega_{\rm peak}&=\frac{2.65\times10^{-6}}{(8\pi)^{1/3}}
 H_{p}R_{*}K_*^2
 \left(\frac{100}{g_{\star\rho}}\right)^{1/3}\Upsilon_{\rm sw},
 \label{eq:acoustic-peak-amplitude}\\
 f_{\rm peak}&=1.9\times10^{-5}\,\mathrm{Hz}\,
 \frac{(8\pi)^{1/3}}{H_{p}R_{*}}
 \frac{T_{\rm red}}{100\,\mathrm{GeV}}
 \left(\frac{g_{\star\rho}}{100}\right)^{1/6},
 \label{eq:acoustic-peak-frequency}
\end{align}
where $h=H_0/(100\,\mathrm{km\,s^{-1}\,Mpc^{-1}})$ is the reduced present Hubble constant. In the sudden-transition approximation with constant vacuum energy, the radiation temperature just after release satisfies $T_{\rm red}^4=30[\rho_r(T_p)+\rho_V]/(\pi^2g_{\star\rho})$. $g_{\star s}=g_{\star\rho}$ and subsequent entropy conservation are assumed. This estimate extrapolates the acoustic fit to strong transitions, assuming rapid conversion into fluid motion without resolving the underlying bubble-wall and plasma dynamics~\cite{Ekstedt:2024fyq,Ai:2024btx}. For a treatment of the energy partition between bubble walls and the plasma, see Refs.~\cite{Ellis:2019oqb,Ai:2023see}.

In our analysis, we use the sound-wave template~\cite{Hindmarsh:2017gnf,Caprini:2019egz}
\begin{equation}
 S_{\rm sw}(x)=x^3\left(\frac7{4+3x^2}\right)^{7/2},\
 n_\Omega(x)=3-\frac{21x^2}{4+3x^2},
 \label{eq:sound-shape-response}
\end{equation}
where $x=f/f_{\rm peak}$. The spectral shape can depend on the transition strength and wall velocity~\cite{Ghosh:2023aum}. Here we neglect its modulation by $\xi$, but allow both the peak amplitude and peak frequency to depend on $\xi$. Eq.~\eqref{eq:spectral-response} therefore includes their logarithmic derivatives with respect to $\xi$. For the strongly supercooled CSI benchmark, $G_\Omega$ and $G_f$ are approximately determined by the response of $H_pR_*$ alone. Since $K_*\to1$ for $\alpha_\theta\gg1$ and $T_{\rm red}$ is fixed by the $\xi$-independent $\rho_V$, $\Omega_{\rm peak}$ and $f_{\rm peak}$ depend on $\xi$ only through $H_{p}R_{*}$, so that with $g\equiv\partial_\xi\ln(H_{p}R_{*})$ and $u_{\rm sw}\equiv H_p\tau_{\rm sw}$,
\begin{equation}
 G_\Omega=\left[1+\frac{u_{\rm sw}(1+2u_{\rm sw})^{-3/2}}{1-(1+2u_{\rm sw})^{-1/2}}\right]g ,
 \quad G_f=-g .
 \label{eq:sudden-responses}
\end{equation}
For the benchmark of Eq.~\eqref{eq:csi-benchmark}, this reduces to $\mathcal T_{\rm src}(f)=0.20\,[5/3+n_\Omega(f)]$, giving $\mathcal T_{\rm src}(f_{\rm peak})\simeq0.33$.

\textbf{Intrinsic anisotropy.} With $P_{X,Y}$ denoting the dimensionless cross spectrum of perturbations $X,Y$, the common source implies, at linear order,
\begin{subequations}
 \begin{align}
 P_{\delta_{\rm GW}^{\rm src}}(f;k)&=\mathcal T_{\rm src}^2(f)P_\zeta(k),\\
 P_{\zeta,\delta_{\rm GW}^{\rm src}}(f;k)&=\mathcal T_{\rm src}(f)P_\zeta(k). \label{eq:source-cross-spectra}
\end{align}   
\end{subequations}
The absolute root-mean-square spectral fluctuation per logarithmic primordial wavenumber is
\begin{equation}
 \Delta_\Omega^{\rm src}(f;k)=\bar\Omega_{\rm GW}(f)
 |\mathcal T_{\rm src}(f)|\sqrt{P_\zeta(k)}.
 \label{eq:source-rms}
\end{equation}

What a detector measures is the angular power spectrum. Writing the total linear response of Eq.~\eqref{eq:source-plus-sw} as $\delta^{\rm tot}_{\rm GW}(f,\hat{\boldsymbol n})=\mathcal T_{\rm tot}(f)\,\zeta(\boldsymbol x_{\mathrm f})$ and using $\int\!\D k\,k^{-1}j_\ell^2(k\Delta\eta)=[2\ell(\ell+1)]^{-1}$ for a scale-invariant $P_\zeta$, with $\ell\ge1$ and $\Delta\eta\equiv\eta_0-\eta_{\mathrm f}$ the conformal propagation interval, one obtains an SW plateau
\begin{equation}
 \frac{\ell(\ell+1)}{2\pi}\,C_\ell^{\rm GW}(f)=\mathcal T_{\rm tot}^2(f)\,P_\zeta ,
 \label{eq:angular-power}
\end{equation}
with the same normalization that reproduces $P_\zeta/25$ for the CMB. 
Here $C_{\ell}^{\mathrm{GW}}$ is the angular power spectrum of $\delta_{\rm GW}^{\rm tot}$.

\textbf{Propagation.} Next we consider anisotropies produced during the propagation of GWs. In the Newtonian gauge, we use
\begin{equation}
 \D s^2=a^2(\eta)\left[-(1+2\Phi_{\rm N})\D\eta^2
 +(1-2\Psi_{\rm N})\D\boldsymbol x^2\right],
 \label{eq:gw-metric-convention}
\end{equation}
with conformal time $\eta$ and scalar potentials $\Phi_{\rm N},\Psi_{\rm N}$. The convention $\zeta=-\Psi_{\rm N}-\mathcal H\delta\rho_{\rm tot}/(\partial_\eta\bar\rho_{\rm tot})$ with $\mathcal H=\partial_\eta\ln a$ agrees with $\zeta=\delta N$. We set the initial conditions for GW propagation after production has ended, assuming a growing superhorizon mode in radiation domination. The GW perturbation in the Newtonian gauge is related to the intrinsic source perturbation on the uniform-density hypersurface by $\delta_{\rm GW}^{\rm N}=\delta_{\rm GW}^{\rm src}+(4-n_\Omega)(\zeta+\Psi_{{\rm N},\mathrm f})$, where $\mathrm f$ denotes the background matching time $\eta_{\mathrm f}$. The factor $4-n_\Omega$ accounts for energy dilution and the shift of frequency bins~\cite{Bartolo:2019oiq,Malhotra:2022ply}.

Place the observer at the origin at conformal time $\eta_0$. The unperturbed ray follows $\boldsymbol x(\eta)=(\eta_0-\eta)\hat{\boldsymbol n}$, with $\boldsymbol x_{\mathrm f}=\boldsymbol x(\eta_{\mathrm f})$. 
Removing the observer contributions to the monopole and dipole and neglecting the gradient-suppressed initial Doppler term, we obtain the observed anisotropy $\delta_{\rm GW}^{\rm tot}(f,\hat{\boldsymbol n})=\delta_{\rm GW}^{\rm src}(f,\boldsymbol x_{\mathrm f})+(4-n_\Omega)\mathcal G_{\rm prop}(\hat{\boldsymbol n})$, where
\begin{align}
 \mathcal G_{\rm prop}(\hat{\boldsymbol n})
 &=\bigl[\zeta+\Psi_{\rm N}+\Phi_{\rm N}\bigr]_{\eta_{\mathrm f},\boldsymbol x_{\mathrm f}}
 \nonumber\\
 &\quad+\int_{\eta_{\mathrm f}}^{\eta_0}\D\eta\,
 \bigl[\partial_\eta\Phi_{\rm N}+\partial_\eta\Psi_{\rm N}\bigr]_{\eta,\boldsymbol x(\eta)}.
 \label{eq:gw-propagation-los}
\end{align}
The term evaluated at $\eta_{\rm f}$ includes the contribution from the change of time slicing, while the integral describes the ISW effect~\cite{Bartolo:2019oiq,Malhotra:2022ply,Cai:2024dya}. For the growing mode in radiation domination with negligible anisotropic stress, $\Phi_{{\rm N},\mathrm f}=\Psi_{{\rm N},\mathrm f}=-2\zeta/3$, so the first term in Eq.~\eqref{eq:gw-propagation-los} is $\zeta+\Psi_{{\rm N},\mathrm f}+\Phi_{{\rm N},\mathrm f}=+\zeta/3-2\zeta/3=-\zeta/3$. The contribution from the change of time slicing, $+\zeta/3$, and the gravitational redshift, $-2\zeta/3$, have opposite signs. The former is separate from the intrinsic PT source.  The modes relevant for us remain superhorizon through radiation--matter equality, during which the potentials evolve to $\Phi_{\rm N}=\Psi_{\rm N}=-3\zeta/5$. The early-ISW integral then contributes $\int\!\D\eta\,[\partial_\eta\Phi_{\rm N}+\partial_\eta\Psi_{\rm N}]=2\left(-\tfrac35+\tfrac23\right)\zeta=+\tfrac{2}{15}\zeta$, and
\begin{equation}
 \mathcal G_{\rm prop}=-\frac{\zeta}{3}+\frac{2\zeta}{15}=-\frac{\zeta}{5},
 \label{eq:gprop-superhorizon-equality}
\end{equation}
which is the matter-era SW result~\cite{Alba:2015cms}. This is the coefficient to be used in the propagation: the multipoles accessible to space-based detectors, $\ell=1,2$, correspond to $k\eta_0\sim\ell$, far below the equality scale $k_{\rm eq}\eta_0=\mathcal O(10^2)$, so these modes are all superhorizon at equality. Combining source and propagation under the same $\zeta=\delta N$ convention gives the total coefficient $\mathcal T_{\rm src}-(4-n_\Omega)/5$ of Eq.~\eqref{eq:source-plus-sw}.

\textbf{SNR of GW anisotropies.} We quantify the GW anisotropy using the ensemble angular-mode SNR, following the definition in Ref.~\cite{LISACosmologyWorkingGroup:2022kbp}, generalized to a covariance that includes the isotropic GW background~\cite{Mentasti:2023icu,Mentasti:2023uyi} and to detector networks:
\begin{align}
\label{eq:snr-1}
\mathrm{SNR}_{\ell}^{2}
&=\int_{0}^{\infty}\D f\,
C_{\ell}^{\mathrm{GW}}(f)
\left[\frac{\bar\Omega_{\mathrm{GW}}(f)}
{\Omega_{\mathrm{GW},\rm noise}^{\ell}(f)}\right]^2,
\\
\mathrm{SNR}^{2}&=\sum_{\ell}\mathrm{SNR}_{\ell}^{2}.
\end{align}
In an orthonormal real spherical-harmonic basis $Y_{\ell m}^{\mathrm R}(\hat{\boldsymbol n})$, these quantities are specified by
\begin{equation}
\label{eq:snr-2}
\begin{aligned}
{\delta^{\rm tot}_{\mathrm{GW}}}(f,\hat{\boldsymbol n})
&=\sum_{\ell,m}\delta_{\ell m}(f)Y_{\ell m}^{\mathrm R}(\hat{\boldsymbol n}),\\
\left\langle\delta_{\ell m}(f)\delta_{\ell'n}(f)\right\rangle
&=\delta_{\ell\ell'}\delta_{mn}C_{\ell}^{\mathrm{GW}}(f),
\end{aligned}
\end{equation}
where $m,n=1,\ldots,2\ell+1$. For $\ell\ge1$, $C_\ell^{\rm GW}$ is given by Eq.~\eqref{eq:angular-power}. Since $\delta^{\rm tot}_{\rm GW}$ has no monopole, $\mathrm{SNR}_0$ is obtained by using $\Omega_{\rm GW}/\bar\Omega_{\rm GW}=1$, i.e.\ $\delta_{00}=\sqrt{4\pi}$ and $C_0^{\rm GW}=4\pi$, in Eq.~\eqref{eq:snr-1}. $\mathrm{SNR}_\ell$ is the expected SNR for a known sky pattern, without component separation.

The effective angular sensitivity $\Omega_{\mathrm{GW},\rm noise}^{\ell}$, including the observing time $T_{\mathrm{obs}}$ ($3\,$yr for the CSI benchmark and $5\,$yr for the benchmark with the mass term, as quoted in the main text) and the statistical fluctuations of the isotropic GW background, is given by
\begin{align}
\label{eq:snr-3}
&\left[\Omega_{\mathrm{GW},\rm noise}^{\ell}(f)\right]^{-2}
=\frac{1}{4\pi}
\left(\frac{3H_0^2}{4\pi^2f^3}\right)^2\nonumber\\
&\qquad\quad\times\sum_{m=1}^{2\ell+1}
\int_{0}^{T_{\mathrm{obs}}}\D t\,
\operatorname{Tr}
\left[
\mathbf P^{-1}\widetilde{\mathbf R}_{\ell m}
\mathbf P^{-1}\widetilde{\mathbf R}_{\ell m}^{\dagger}
\right],
\end{align}
where $\mathbf P(f,t)=\mathbf N(f)+\mathbf P_{\mathrm{iso}}(f,t)$ is the full one-sided spectral covariance matrix, evaluated for the fiducial isotropic background. The trace runs over the detector channels, and the inverse covariance weights the angular responses while accounting for all channel correlations. For $\mathbf P=\mathbf N$ and a time-independent response, $\Omega_{\mathrm{GW},\rm noise}^{\ell}$ reduces to the sensitivity of Ref.~\cite{LISACosmologyWorkingGroup:2022kbp} divided by $\sqrt{T_{\rm obs}}$. We use the instrumental noise models of Refs.~\cite{LISACosmologyWorkingGroup:2022kbp,Wang:2021njt,Ruan:2020smc} to construct $\mathbf N$, which is rescaled together with the response by $4(f/f_*)^2|W(f)|^2$, as in Ref.~\cite{LISACosmologyWorkingGroup:2022kbp}, so that the time-delay interferometry (TDI) generation drops out. Here $f_*=c/(2\pi L)$, with $c$ the speed of light, and $W(f)$ is the TDI delay factor: $|W(f)|^2=1$ for TDI 1.0 and $4\sin^2(f/f_*)$ for TDI 1.5 in the convention of Ref.~\cite{LISACosmologyWorkingGroup:2022kbp}.
We take the optical-metrology-system (OMS) noise amplitude $P=15$, the test-mass acceleration noise amplitude $A=3$ and the arm length $L=2.5\times10^{9}\,\mathrm{m}$ for LISA, and $P=8$, $A=3$, $L=3\times10^{9}\,\mathrm{m}$ for Taiji; the dimensionless $P$ and $A$ normalize the OMS displacement and test-mass acceleration amplitude spectral densities to $P\,\mathrm{pm}/\sqrt{\mathrm{Hz}}$ and $A\,\mathrm{fm}\,\mathrm{s}^{-2}/\sqrt{\mathrm{Hz}}$. We use the standard TDI weights and uncorrelated noise between channels and detectors. 

The isotropic GW covariance and angular response are
\begin{subequations}
\label{eq:snr-4}
\begin{align}
\mathbf P_{\mathrm{iso}}(f,t)
&=\frac{3H_0^2}{4\pi^2 f^3}
\bar\Omega_{\mathrm{GW}}(f)\,
\widetilde{\mathbf R}_{00}(f,t),\\
\widetilde R_{\ell m,IJ}(f,t)
&\equiv\frac{1}{8\pi}\int \D \hat{\boldsymbol n}\,
\widetilde Y_{\ell m}(\hat{\boldsymbol n})
\nonumber\\
&\quad\times\sum_{\lambda=+,\times}
E_I^\lambda(f,\hat{\boldsymbol n},t)
E_J^{\lambda *}(f,\hat{\boldsymbol n},t).
\end{align}
\end{subequations}
Here $\widetilde Y_{\ell m}=\sqrt{4\pi}\,Y_{\ell m}^{\mathrm R}$; the monopole is denoted separately by $\widetilde Y_{00}=1$. The quantity $E_I^\lambda(f,\hat{\boldsymbol n},t)$ is the reduced response of channel $I$ to a GW of polarization $\lambda=+,\times$ from direction $\hat{\boldsymbol n}$ (polarization tensors normalized as $e^\lambda_{ab}e^{\lambda'*}_{ab}=\delta_{\lambda\lambda'}$), including the geometrical propagation phases and, for LISA--Taiji, the detector separation. The indices $I,J$ label the $A,E,T$ TDI channels, constructed from time-delayed laser measurements to suppress laser frequency noise. For the LISA--Taiji network, these indices also identify the constellation. All response functions and noise spectra use the same channel normalization. The response matrices are dimensionless, while $\mathbf P$ and $\mathbf P_{\mathrm{iso}}$ have units of $\mathrm{Hz}^{-1}$. With the observing time included as above, $\Omega_{\mathrm{GW},\rm noise}^{\ell}$ has units of $\mathrm{Hz}^{1/2}$. We use the full three-channel covariance for LISA and six-channel covariance for the LISA--Taiji network, including the GW cross-correlations between the two constellations. For LISA alone, the sum over $m$ of the trace is independent of $t$, so that $\int_0^{T_{\rm obs}}\D t=T_{\rm obs}$; for LISA--Taiji it is averaged over one orbital period.

In the noise-dominated limit with independent instrumental channel noises, \(P_I\simeq N_I\) and the trace reduces to the usual weighting by \(1/(N_I N_J)\)~\cite{LISACosmologyWorkingGroup:2022kbp}. The benchmarks considered here are instead dominated by the monopole, so our forecasts account for monopole self-noise through \(\mathbf P_{\mathrm{iso}}\)~\cite{Mentasti:2023icu,Mentasti:2023uyi}: the weights are \(1/(P_IP_J)\) with \(P_I=N_I+P_{\mathrm{iso},II}\) for LISA, and the full \(6\times6\) inverse for the network. The frequency integral uses positive frequencies, and the sum over harmonic modes already includes all \(2\ell+1\) modes.

\section{Effective potential and conditions for a frozen spectator}
\label{app:csi-consistency}

\textbf{Potential and matching.} The effective potential, including the gauge Coleman--Weinberg potential in Eq.~\eqref{eq:csi-cw-potential}, the finite-temperature gauge-boson contribution, and the fermion one-loop vacuum and finite-temperature contributions, is~\cite{Dolan:1973qd,Carrington:1991hz}
\begin{equation}
 V_{\rm eff}=V_{\rm CW}^{(g)}+\Delta V_0^{(\psi)}
 +\Delta V_T^{(V)}+\Delta V_T^{(\psi)}.
 \label{eq:csi-full-potential}
\end{equation}
At renormalization scale $Q=v_\phi$, these contributions are
\begin{align}
 \Delta V_0^{(\psi)}=&-\frac{n_\psi m_\psi^4}{64\pi^2}
 \left(\ln\frac{m_\psi^2}{Q^2}-\frac32\right),
 \label{eq:csi-fermion-vacuum}\\
 \Delta V_T^{(\psi)}=&\frac{n_\psi T^4}{2\pi^2}
 J_F\!\left(\frac{m_\psi^2}{T^2}\right),
 \label{eq:csi-fermion-thermal}\\
 \Delta V_T^{(V)}=&\frac{3T^4}{2\pi^2}
 J_B\!\left(\frac{g_D^2\phi^2}{T^2}\right)
 \nonumber\\
 &+\frac{T}{12\pi}
 \left[g_D^3 \phi^3-(g_D^2\phi^2+\Pi_V)^{3/2}\right],
 \label{eq:csi-vector-thermal}
\end{align}
where $n_\psi=4$ for one Dirac fermion and $\Pi_V=g_D^2T^2/3$ is the vector Debye mass squared in the adopted approximation. Our thermal-function convention is
\begin{equation}
 J_{B,F}(y)=\pm\int_0^\infty \D z\,z^2
 \ln\!\left(1\mp \e^{-\sqrt{z^2+y}}\right),
 \label{eq:csi-thermal-functions}
\end{equation}
with the upper signs for bosons and lower signs for fermions. 
We neglect scalar and Goldstone thermal loops and use a canonical radial kinetic term, assuming the other scalar directions do not participate in tunneling.

The mass profile $m_\psi^2=\xi^2 v_\phi^2 F(\phi)$ is used on $0\leq\phi\leq v_\phi$ and continued by $F=0$ for $\phi\geq v_\phi$. Since $F$ vanishes at both vacua, $\Delta V_0^{(\psi)}$ and $\partial_\phi\Delta V_0^{(\psi)}$ also vanish there. No additional finite quartic or constant counterterm is needed to restore the zero-temperature vacuum. Other finite higher-dimensional potential coefficients are set to zero at $Q$, and all matching coefficients are held fixed as $\xi$ varies. This specifies an effective model, without asserting a renormalizable ultraviolet completion or all-order protection of the spectator.

\textbf{Spectator evolution.} Let $U_{\rm eff}(\sigma,N)$ denote the effective homogeneous spectator potential including vacuum and thermal corrections, with the transition field at its local equilibrium value in each phase. The effective mass of the spectator satisfies $m_{\sigma,\rm eff}^2=\partial_\sigma^2U_{\rm eff}$. When spectator decay is neglected, its equation of motion is
\begin{equation}
 \sigma''+\left(3+\frac{H'}H\right)\sigma'
 +\frac{\partial_\sigma U_{\rm eff}}{H^2}=0.
 \label{eq:csi-spectator-evolution}
\end{equation}
A small instantaneous mass does not by itself guarantee negligible accumulated motion. We therefore estimate the accumulated displacement of the spectator. Near the symmetric phase, $m_\psi^2\simeq\sigma^2\phi^2/\Lambda^2$. Combining the leading fermion thermal correction $\Delta V_T^{(\psi)}\simeq m_\psi^2T^2/12$ with the relativistic radial fluctuation $\langle\phi^2\rangle_T\simeq T^2/12$ gives
\begin{equation}
 \Delta U_{\rm th}\simeq\frac{\sigma^2T^4}{144\Lambda^2},\quad
 m_{\sigma,\rm th}^2\simeq\frac{T^4}{72\Lambda^2},
 \label{eq:csi-residual-thermal-mass}
\end{equation}
which corresponds to the two-loop correction at $\phi=0$. With only this thermal contribution included, the homogeneous spectator obeys
\begin{equation}
 \ddot\sigma+3H\dot\sigma+m_{\sigma,\rm th}^2\sigma=0,
 \label{eq:csi-thermal-drift}
\end{equation}
where dots denote proper-time derivatives. The positive thermal mass drives $\sigma$ toward the origin even if initially at rest, producing an accumulated displacement $\Delta\sigma(N)=\sigma(N)-\sigma(N_{\mathrm i})$.

In vacuum domination, $m_{\sigma,\rm th}^2/H^2$ falls as $a^{-4}$. For Eq.~\eqref{eq:csi-benchmark}, evolving the spectator under this thermal correction from $T_c$ with $\sigma'=0$ results in $|\Delta\sigma|/|\bar\sigma|\simeq5\times10^{-4}$ before percolation. After reheating, let $T$ denote the radiation temperature. 
Expanding the polynomial mass profile about $\phi=v_\phi$ from below gives $m_\psi^2\simeq4\xi^2(\phi-v_\phi)^2$. Neglecting Boltzmann suppression of scalar fluctuations, we obtain $m_{\sigma,\rm th}^2\lesssim4T^4/(72\Lambda^2)$ and the total fractional drift $|\Delta\sigma|/|\bar\sigma|\lesssim4.4\times10^{-3}$ by the time the density reaches $\rho_{\mathrm f}=10^{-4}\rho_V$. This small thermal drift supports treating the homogeneous spectator as approximately constant over this interval.

\bibliographystyle{utphys}
\bibliography{Ref}

\end{document}